\pdfoutput=1  % ensure pdflatex
\documentclass[fleqn,11pt]{wlscirep}
\usepackage[utf8]{inputenc}
\usepackage[T1]{fontenc}
\usepackage{csquotes}
\usepackage{graphicx}
\usepackage{subcaption}
\usepackage{booktabs, tabularx}
\usepackage{amsmath}
\usepackage{mathrsfs}
\usepackage[export]{adjustbox}
\usepackage{pgf, tikz, pgfplots}
\usetikzlibrary{arrows,shapes,positioning, fadings}
\pgfplotsset{compat=1.17}
\usepackage{hyperref}
\usepackage{lineno}

\title{Enhanced ultrafast X-ray diffraction by transient resonances}

\author[1,2,3*]{Stephan Kuschel}
\author[4 *]{Phay J. Ho}
\author[4,5]{Andre Al Haddad}
\author[2,3,6]{Felix Zimmermann}
\author[7]{Leonie Flueckiger}
\author[2,3]{Matthew R. Ware}
\author[3]{Joseph Duris}
\author[3]{James P. MacArthur}
\author[3]{Alberto Lutman}
\author[3]{Ming-Fu Lin}
\author[3,8]{Xiang Li}
\author[3]{Kazutaka Nakahara}
\author[3]{Jeff W. Aldrich}
\author[3]{Peter~Walter}
\author[4,9]{Linda~Young}
\author[4,5,10]{Christoph~Bostedt}
\author[3*]{Agostino Marinelli}
\author[1,2,3*]{Tais Gorkhover}

\affil[1]{University of Hamburg, Institute for Experimental Physics/CFEL,
Luruper Chaussee 149
22761 Hamburg, Germany}
\affil[2]{Stanford PULSE Institute, SLAC National Accelerator Laboratory, 2575 Sand Hill Rd, Menlo Park, CA 94025, USA}
\affil[3]{SLAC National Accelerator Laboratory, 2575 Sand Hill Rd, Menlo Park, CA 94025, USA}
\affil[4]{Chemical Sciences and Engineering Division, Argonne National Laboratory, 9700 S. Cass Avenue, Lemont, IL 60439, USA}
\affil[5]{Paul Scherrer Institute, 5232 Villigen, Switzerland}
\affil[6]{IOAP, TU Berlin, Hardenbergstrasse 36, 10623 Berlin }

\affil[7]{La Trobe University, 1300 La Trobe, Australia}
\affil[8]{J.R. Macdonald Laboratory, Department of Physics, Kansas State University, Manhattan, Kansas 66506, USA}
\affil[9]{Department of Physics and James Franck Institute, The University of Chicago, Chicago, IL 60637, USA}
\affil[10]{Ecole Polytechnique Fédérale de Lausanne (EPFL),1015 Lausanne, Switzerland}
\affil[*]{kuschel@slac.stanford.edu}
\affil[*]{tais.gorkhover@cfel.de}
\affil[*]{pho@anl.gov}
\affil[*]{marinelli@slac.stanford.edu}

\definecolor{royalblue}{HTML}{4169E1}  % 5fs arount M edge color

\newcommand{\suppref}[1]{(see supplement, section #1)}

\begin{document}

%%%%%%%%%%%%%% Introductory paragraph %%%%%%%%%%%%%%%%%%%%%%%
%  200 words max. and fully referenced.
\begin{abstract}
Diffraction-before-destruction imaging with single ultrashort X-ray pulses has the potential to visualise non-equilibrium processes, such as chemical reactions, at the nanoscale with sub-femtosecond resolution in the native environment without the need of crystallization\cite{Neutze2000,chapman2007,Bostedt2010,Bostedt2012,Gorkhover2016,Seibert2011}. Here, a nanospecimen partially diffracts a single X-ray flash before sample damage occurs. The structural information of the sample can be reconstructed from the coherent X-ray interference image. State-of-art spatial resolution of such snapshots from individual heavy element nanoparticles is limited to a few nanometers\cite{Ayyer2021,Gorkhover2016, Gorkhover2018, Aquila2015}. Further improvement of spatial resolution requires higher image brightness which is ultimately limited by bleaching effects\cite{Young2010,Son2011,Ho-2016-PRA} of the sample. We compared snapshots from individual 100\,nm Xe nanoparticles as a function of the X-ray pulse duration and incoming X-ray intensity in the vicinity of the Xe M-shell resonance. Surprisingly, images recorded with few femtosecond and sub-femtosecond pulses are up to 10 times brighter than the static linear model predicts \cite{Guinier1955,Attwood2012, Deslattes1968}. Our Monte-Carlo simulation and statistical analysis of the entire data set confirms these findings and attributes the effect to transient resonances. Our simulation suggests that ultrafast form factor changes during the exposure can increase the brightness of X-ray images by several orders of magnitude. Our study guides the way towards imaging with unprecedented combination of spatial and temporal resolution at the nanoscale.
\end{abstract}

\maketitle

%%%%%%%%%%%%%% Motivation 400 words%%%%%%%%%%%%%%%%%%%%%%%%%%
 %\textbf{Text}

%\linenumbers
The combination of coherent diffraction imaging (CDI) with bright and focused femtosecond short flashes from X-ray Free Electron lasers (FELs) has brought to light transient phenomena such as metastable stages of metal nanoparticle formation\cite{Barke2015}, unexpected morphologies diversity  in soot formation\cite{Loh2012}, complex light matter interaction dynamics\cite{Bostedt2012, Bostedt2010,Gorkhover2012,Rupp2020,Ferguson2016,Gorkhover2012, Ihm2019direct}, vortex organisation in superfluid droplets\cite{Gomez2014}, relaxation of superheated particles on the femtosecond time scale \cite{Gorkhover2012, Gorkhover2016, Peltz2022few} and otherwise inaccessible states of water\cite{Sellberg2014}.

Theoretical studies suggest that sub-nm spatial resolution within a single FEL exposure is feasible under ideal FEL pulse conditions \cite{Ho-2016-PRA, son2011impact, Schropp2010,Neutze2000, Aquila2015}. However, the state-of-art spatial resolution in experimental X-ray CDI from an individual heavy element nanoparticle is limited to a few nanometers \cite{Ayyer2021}. Further increase in resolution requires an enhancement in diffraction image brightness, which is proportional to the incoming X-ray photon intensity $I_{ph}$ and the wavelength dependent elastic scattering cross section $\sigma_{scat}$ of the material \cite{Guinier1955,Attwood2012, Howellst1995}. The image brightness does not scale linearly with $I_{ph}$ for very high photon intensities due to bleaching effects of the sample which can decrease the effective scattering cross section during the FEL exposure \cite{Young2010,Son2011,Aquila2015, Hau-Riege2008,Barty2011,Ho2020Sucrose,Ayyer2021, Ferguson2016,Schorb2012,Ho-2017-JPB}.

The bleaching is a result of a complex interplay between electronic and ionic structural damage of the sample interacting with an intense X-ray FEL pulse. Electronic damage is initiated by photoionisation processes. Sequential multi-photon absorption and subsequent relaxation processes such as Auger/Coster-Kronig decays remove the electrons from the parent ions on sub-fs to 10-fs time scale. Electron delocalisation increases the transparency of the sample as a shrinking number of bound electrons usually decreases the absorption/scattering cross sections of an ion\cite{Young2010,Son2011, Hau-Riege2008}. The subsequent damage to the ionic structure through hydrodynamic expansion is driven by the hot electrons trapped by a rapidly build up space charge. Overall, most experimental and theoretical studies suggest that photoionisation is  detrimental to the quality of the the images \cite{Aquila2015, Hau-Riege2008,Barty2011,Ho2020Sucrose,Ayyer2021, Ho-2017-JPB, son2011impact, Ayyer2021}.

In contrast to previous studies, we find that photoionisation can increase the effective scattering cross section by at least one order of magnitude if FEL pulses are extremely short and tailored to excite transient resonances (TRs). TRs arise when core-hole ionisation events drive the parent ion into  short-lived resonances \cite{Kanter2011, Rudek2012a,Bostedt2010,Bostedt2012,Rupp2020,Ho2020Sucrose}. Previous studies in rare gases reported increases in absorption cross section by almost two orders of magnitude due to individual transient resonances in Ne\cite{Kanter2011} and even transient resonances cascades in Xe atoms at 1500\,eV\cite{Rudek2012a}. The potential of transient resonances for imaging received little attention so far as first experiments with 100-200 fs long pulses on single nanoparticles suggested detrimental effects to diffraction brightness due to accelerated sample decomposition or nanoplasma effects\cite{Ho2020Sucrose,Bostedt2012,Bostedt2010,Rupp2020}.

\begin{figure}[b]
  \centering
  \textbf{a)}\quad
  \includegraphics[width=0.68\textwidth, valign=t]{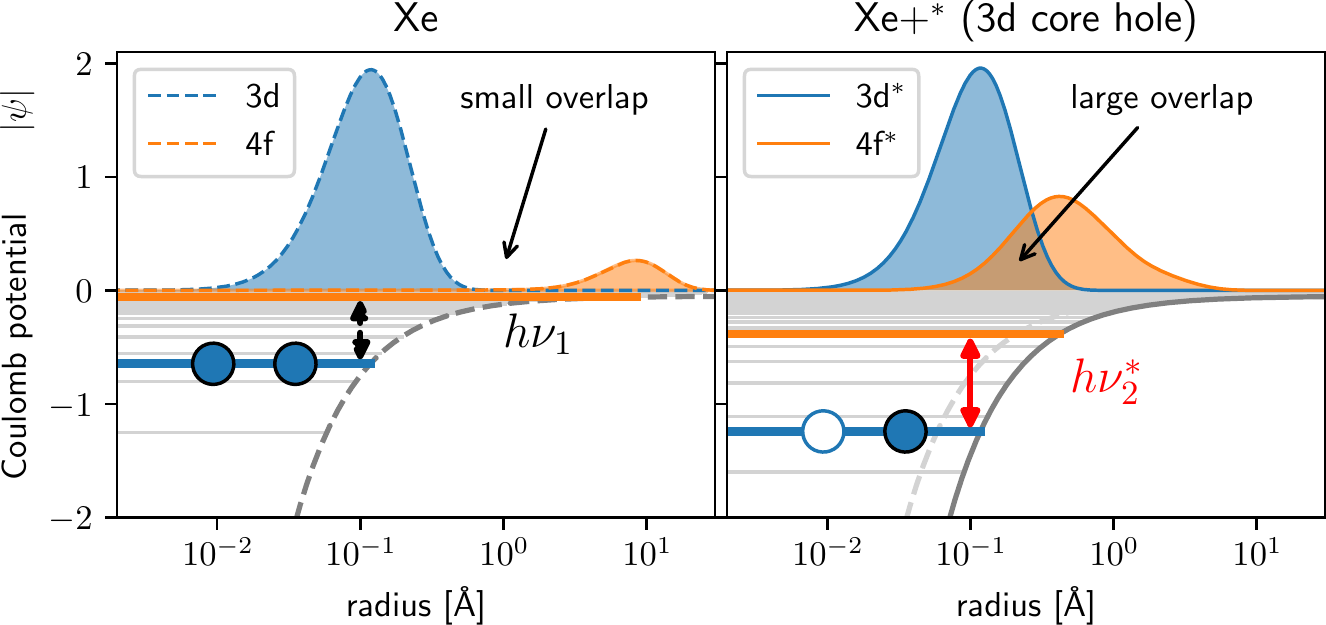} \\[1.5em]
  \textbf{b)}\quad
  \includegraphics[width=0.48\textwidth, valign=t]{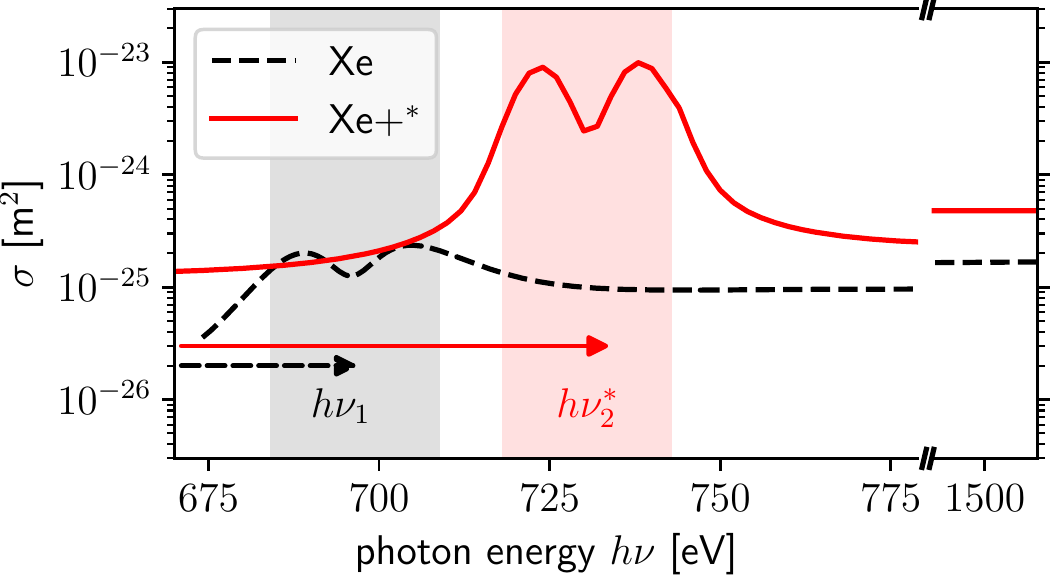}

  \caption{The concept of transient resonances: In (a, top part), the Coulomb potential is plotted versus the distance from the charged core for a neutral Xe atom (gray dotted line, right side) and for a core excited Xe+$^*$ (gray solid line, left side). The corresponding 3d (blue) and 4f orbitals (orange)  are displayed for a neutral Xe atom (left, dashed lines) and 3d$^*$ and 4f$^*$ excited Xe ion with a core-hole (right, solid lines). First, the core-hole excitation shifts the binding energy of the remaining 3d$^*$ electrons from $h\nu_1$ (black) to greater $h\nu_2^*$ (red) due to a modified Coulomb potential (a, bottom, right, solid grey line). Second, the core-hole excitation also rearranges the orbitals to 3d$^*$ (blue) and 4f$^*$ (orange). Due to the core-hole induced charge imbalance the 4f$^*$ orbital is pulled closer to the atom's center by almost two orders of magnitude creating a strongly increased overlap with the 3d$^*$ wave function and hence an increased transition dipole strength for the 3d$^*~\rightarrow~\text{4f}^*$ transition. This has dramatic consequences for the scattering cross section in the vicinity of the Xe 3d absorption edge. In (b), the scattering cross section $\sigma_{scat}$ for the neutral Xe (dashed black line) and $\sigma_{scat}^*$ for the excited Xe+$^*$ (red solid line) is plotted vs the incoming X-ray photon energy. The excited atom Xe$+^*$ scatters two orders more strongly than neutral Xe at the shifted resonance position $h\nu_2^*$ and almost two orders more compared to the neutral Xe resonance maximum $h\nu_1$. Energies, cross-sections and radial wave functions displayed were calculated using Hartree-Fock simulations \suppref{Simulations}.
    }
  \label{fig:theoryTR}
\end{figure}

The TR mechanism is illustrated in \autoref{fig:theoryTR} based on the mechanism of 3d~$\rightarrow$~4f resonant scattering in a neutral Xe atom and a core-excited Xe+$^*$ ion. Resonant scattering occurs through virtual electron transitions between the 3d core level and the unoccupied 4f state as illustrated for neutral Xe ( \autoref{fig:theoryTR} a, left panel). If X-rays are scattered by the core-hole excited Xe+$^*$ ion with a modified Coulomb potential ( \autoref{fig:theoryTR} a, right panel), the binding energy of the excited 3d$^*$ orbital shifts from $h\nu_1$ (black dashed) to $h\nu_2^* \approx h\nu_1+40 $\,eV (red). In addition, the 4f$^*$ orbital in Xe+$^*$  is pulled closer towards the core in order to screen the positive charge visualised as the absolute value of the wave function $|\psi|$ ( \autoref{fig:theoryTR} a, top of left panel). This orbital restructuring increases the overlap between the 3d$^*$ and 4f$^*$ wavefunctions. The resulting 3d$^*$~$\rightarrow$~4f$^*$ transition matrix element   augments the resonance scattering cross section $\sigma_{scat}^*$ by more than one order of magnitude above the neutral Xe $\sigma_{scat}$(red vs dotted line in \autoref{fig:theoryTR}b; see Supplements for detailed descriptions of the calculation). The overall lifetime of a single transient resonance is limited by the core-hole decay time or photoionisation rates. A previous absorption spectroscopy study on FEL excited Xe atoms suggested that sequential multiphoton absorption can trigger TR cascades throughout the entire FEL exposure duration at 1500\,eV \cite{Rudek2012a}.

%%%%%%%%%%%%%%%%%%%%%%%%%%Figure 2 400 words%%%%%%%%%%%%%%%%%%%%%%%%%%
In our study, we explore how TRs and TR cascades can influence the brightness and spatial resolution of diffraction images of individual Xe nanoparticles as a function of of FEL pulse parameters such as FEL photon energy $h\nu$, pulse energy $E_\mathrm{p}$ and duration $\tau$ (all parameters are summarized in Table 1, Supplements).

A detailed schematic of our experiment is exhibited in Supplements. Individual near-spherical Xe nanoparticles with diameters 40-150\,nanometers intersect the path of focused and intense single X-ray FEL pulses inside the LAMP endstation at the Linac Coherent Light Source (LCLS)\cite{Ferguson2015, Emma2010}. Single-particle, single-exposure X-ray diffraction snapshots are recorded at 120\,Hz using p-n junction charge coupled devices (pnCCDs)\cite{Strueder2010} located at two different positions further downstream from the interaction region. The particle size is directly encoded into the Airy pattern-like diffraction patterns and can be recovered with $\pm 0.3$\,nm precision  \suppref{Cluster size fitting model}.
We scanned the photon energy $h\nu$ in the vicinity of the Xe 3d absorption edge between $650\,\mathrm{eV} < h\nu < 740\,\mathrm{eV}$. In addition, we recorded images at $h\nu=1500\,\mathrm{eV}$ based on previously observed TR cascades in Xe ion spectra\cite{Rudek2012}. Each scanning step contains thousands of diffraction patterns from single Xe nanoparticles with fluctuating brightness due to random positions inside the 1.5\,$\mu$m FEL focus with a near Gaussian X-ray intensity distribution \cite{Gorkhover2012} .  We investigated three FEL pulse durations: 200\,fs as a typical pulse duration preferred in most ultrafast CDI studies \cite{Hantke2014,Loh2012, Seibert2011, Aquila2015, Gorkhover2018, Daurer2019}, 5-10\,fs as a pulse duration beating the onset of the ionic structure damage of the specimen, and newly available sub-fs pulses \cite{Duris2020} outrunning the electronic damage through Auger decay (Xe M-shell Auger life time $\approx$1 fs). The FEL pulse energy increased in general with the pulse duration, we made measurements with attenuated 200-fs FEL pulse duration as a reference.

\begin{figure}
    \centering
    \begin{tikzpicture}
    \definecolor{pulse}{RGB}{0,51,204}
    \node at (0,-1.4) {};  % maintain baseline
    \fill[color=pulse] plot[domain=-0.5:0.5, samples=20, smooth] (\x, {0.3*exp(-500*\x*\x)) + 6.5}) -- cycle;
    \draw[->, -triangle 45,color=pulse] (-0.3, 6.5) -- (0.7,6.5);
    \node[align=left] at (0.9, +0.8 + 6.5) {\textbf{<1\,fs}};
    %\fill[color=pulse, opacity=0.3, overlay, path fading=east] (1.3, 6.5) -- (2.9, 8) -- (2.9, 5.3) -- cycle;
    \node at (1.3, 6.5) {\includegraphics[width=0.6cm]{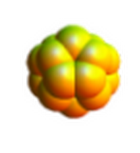}};

    \fill[color=pulse] plot[domain=-0.9:0.9, samples=20, smooth] (\x, {1.5*exp(-12*\x*\x)) + 3.5}) -- cycle;
    \draw[->, color=pulse, -triangle 45] (-0.7, 3.5) -- (1.0, 3.5);
    \node[align=left, fill=white, fill opacity=0.0, text opacity=1] at (0.9, +0.8 + 3.5) {\textbf{200\,fs}};
    \fill[color=pulse, opacity=0.3, overlay, path fading=east] (1.3, 3.5) -- (2.9, 5) -- (2.9, 2.3) -- cycle;
    \node at (1.3, 3.5) {\includegraphics[width=0.6cm]{img/cluster.png}};
    \node[anchor=north] at (0,3.5) {XFEL};

    \fill[color=pulse] plot[domain=-0.5:0.5, samples=20, smooth] (\x, {0.5*exp(-70*\x*\x)) + 0.5}) -- cycle;
    \draw[->, color=pulse, -triangle 45] (-0.4, 0.5) -- (0.7, 0.5);
    \node[align=left, fill=white, fill opacity=0.75, text opacity=1] at (.9, 0.8 + 0.5) {\textbf{5\,fs}};
    %\fill[color=pulse, opacity=0.3, overlay, path fading=east] (1.3, .5) -- (2.9, 2) -- (2.9, -0.7) -- cycle;
    \node at (1.3, 0.5) {\includegraphics[width=0.6cm]{img/cluster.png}};
    \end{tikzpicture}
    \includegraphics[width=14cm]{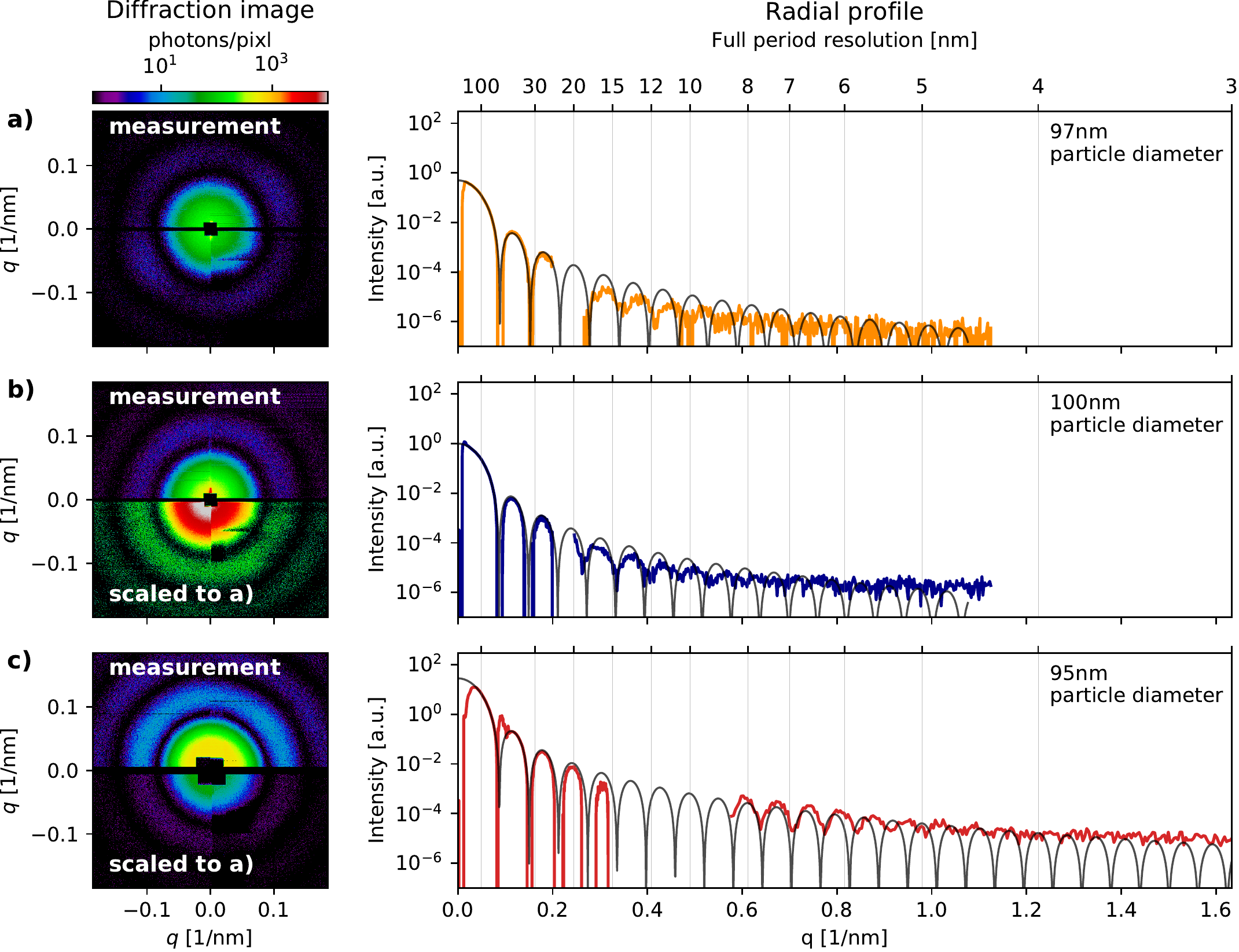}

    \caption{The brightest X-ray diffraction patterns of individual Xe nanoparticles with diameters 97$\pm$3\,nm are displayed for sub-fs (a), 200\,fs (b) and 5-10\,fs (c) pulse durations on the left side in the top part of the square. The  sub-fs image was recorded with a $E_{p}=0.02$\,mJ pulse energy at $h\nu=735$\,eV X-ray photon energy, the 200\,fs with  $E_{p}=1.5$\,mJ at $h\nu=730$\,eV, the 5\,fs  $E_{p}= 0.1$\,mJ pulse energy at $h\nu=1500$\,eV. The corresponding one dimensional radial profiles vs the scattering wave vector values $q$ are shown on the right side. The measured radial profiles are compared with particle size-fitted diffraction patterns of ideal spheres (grey). The brightness of the 200\,fs and 5\,fs diffraction patterns are normalized by the the incoming pulse energies difference relative to the sub-fs snapshot and displayed in the bottom part of the of the square. According to linear scaling with incoming X-ray pulse energy one expects that the brightest X-ray diffraction snapshots can be observed from 1.5\,mJ, 200\,fs FEL pulses ((b),dark blue). In contrast, we find the the diffraction patterns which were recorded with only 0.1\,mJ and 5\,fs short pulses are the absolut brightest. We attribute the increased brightness to transient resonances. Diffraction signal levels down to 5\,nm resolution are visible despite noise from fluorescence in (c) and the overall radial profile follows closely the theoretical calculation for diffraction from a solid sphere.
    \cite{Hau-Riege2008, Bostedt2012, Rupp2020}. See \suppref{Single exposure images} for the corresponding full detector images.
    }
    \label{fig:experiment}
\end{figure}

The comparison of the brightest images for each pulse duration sheds light on the scattering efficiency and the role of TRs at different time scales. The brightest single X-ray diffraction snapshots per FEL pulse duration are displayed on the left side of \autoref{fig:experiment}.The diffraction patterns stem from similarly sized nanoparticles with diameters 97$\pm$3\,nm. Intuitively, one would expect that the brightest images are recorded using 200\,fs pulse with the highest pulse energy $E_{p}=1.5\,\mathrm{mJ}$ which corresponds to the highest $I_{ph}$ (panel b). In fact, the most intense X-ray diffraction snapshots were taken at $h\nu=1500\,\mathrm{eV}$ with few femtosecond pulses and only $E_{p}=0.12\,\mathrm{mJ}$ (panel c), (\autoref{fig:experiment}). The sub-fs FEL pulse duration outruns most damage mechanisms and thus, serves as reference measurement recorded with $E_{p}=0.02\,\mathrm{mJ}$ (panel a). The brightness of the lower half of the images is linearly scaled by their respective pulse energies relative to the sub-fs FEL reference . The 200-fs exposure is 50 times dimmer than linear scaling predicts which indicates detrimental bleaching. Surprisingly, the 5-fs snapshot is almost 10\,times brighter than the normalized sub-fs image which suggests a non-linear increase in scattering efficiency.

%%%%%%%%%%%%%%%%%%%%%%%%%%%%%%Figure 600 words%%%%%%%%%%%%%%%%%%%%%%%%%%

\begin{figure}
   \centering
    \includegraphics[height=9cm]{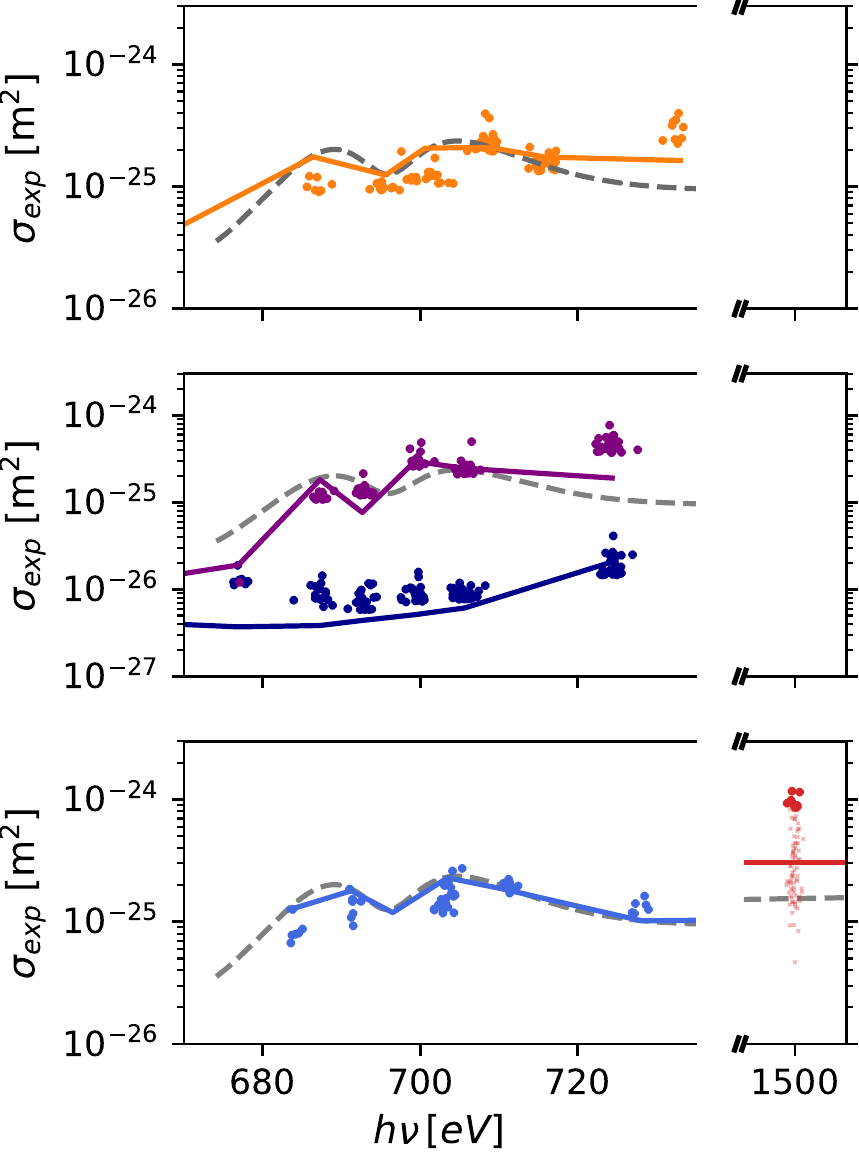}
    \begin{tikzpicture}[overlay]
    \node[color=orange!90!black] at (-4.3, 8.6) {$E_{p}=0.03\mathrm{mJ}$};
    \node[color=purple!50!black] at (-4.3,5.8) {$E_{p}=0.02\mathrm{mJ}$};
    \node[color=blue!50!black] at (-1.7, 4.0) {$E_{p}=1.5\mathrm{mJ}$};
    \node[color=royalblue!90!black] at (-4.2, 2.9) {$E_{p}=0.01\mathrm{mJ}$};
    \node[color=red!80!black] at (-1.7, 2.9) {$E_{p}=0.1\mathrm{mJ}$};
    \end{tikzpicture}
    \begin{tikzpicture}
    \definecolor{pulse}{RGB}{0,51,204}
    \node at (0,-1.4) {};  % maintain baseline
    \fill[color=pulse] plot[domain=-0.5:0.5, samples=20, smooth] (\x, {0.5*exp(-250*\x*\x)) + 6.2}) -- cycle;
    \node[align=center] at (0, 6.2 -0.3) {<1\,fs};

    \fill[color=pulse] plot[domain=-1:1, samples=20, smooth] (\x, {1.5*exp(-12*\x*\x)) + 3.0}) -- cycle;
    \node[align=center, fill=white, fill opacity=0.0, text opacity=1] at (0, + 3.0 - 0.3) {200\,fs};

    \fill[color=pulse] plot[domain=-0.5:0.5, samples=20, smooth] (\x, {0.5*exp(-70*\x*\x)) + 0.3}) -- cycle;
    \node[align=center, text opacity=1] at (0, 0.3 -0.3) {5\,fs};
    \end{tikzpicture}
    \includegraphics[height=9cm]{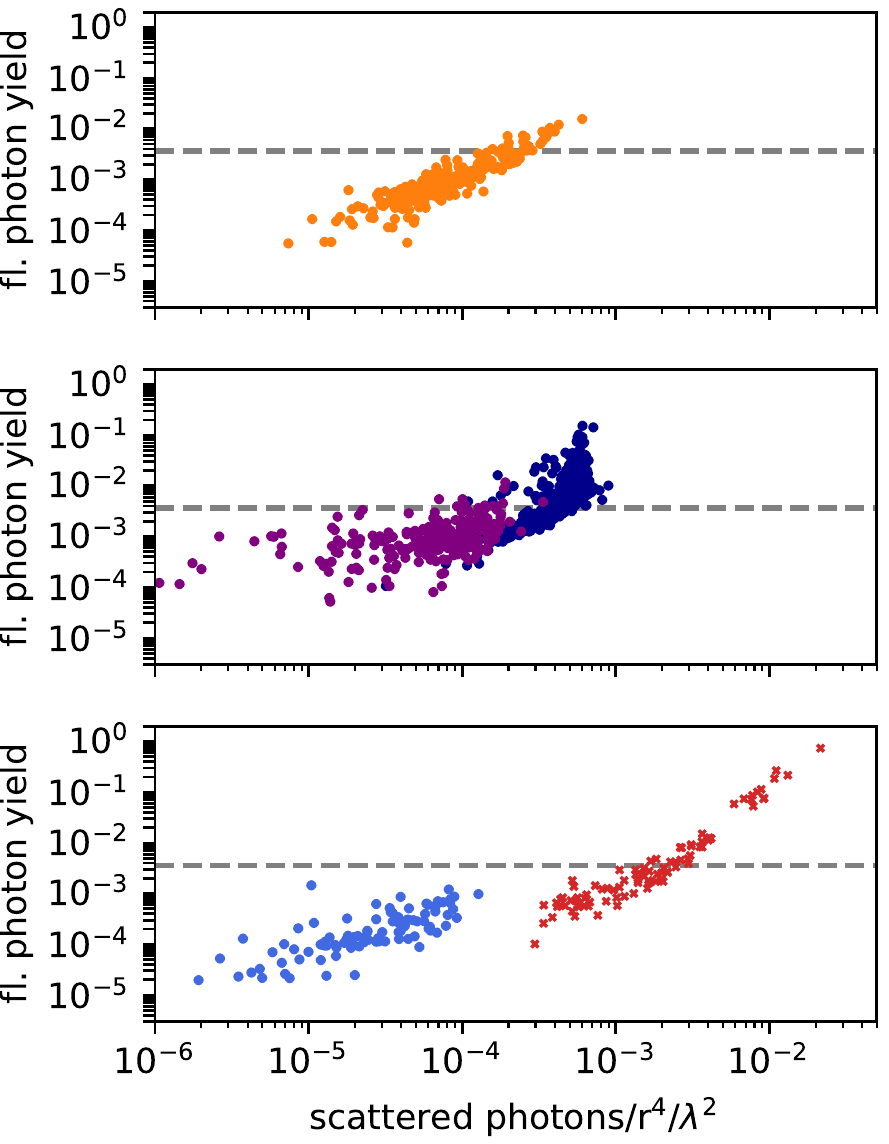}
    \begin{tikzpicture}[overlay]
    \node[color=orange!90!black] at (-4.8, 8.6) {$h\nu=735\,\mathrm{eV}$};
    \node[color=purple!50!black] at (-4.8, 5.8) {$h\nu=730\,\mathrm{eV}$};
    \node[color=blue!50!black] at (-1.9, 4.5) {$h\nu=730\,\mathrm{eV}$};
    \node[color=royalblue!90!black] at (-4.8, 3) {$h\nu=730\,\mathrm{eV}$};
    \node[color=red!80!black] at (-1.5, 1.3) {$h\nu=1500\,\mathrm{eV}$};
    \end{tikzpicture}
    \caption{The left side depicts the measured $\sigma_{exp}$ in dependence to the incoming X-ray photon energy $h\nu$ for all pulse duration measurements with average pulse energy $E_p$. The gray dashed line indicates the neutral Xe $\sigma$ for comparison. Our calculations (solid lines) agree overall well with the experiment. The strongest enhancement can be found at 1500\,eV (red) where TR cascades have been observed in the past in Xe atoms \cite{Rudek2012}. We analysed all photon energy scan steps with enhanced $\sigma_{exp}$ as a correlation between scattering and absorption efficiency (represented by fluorescence yield per atom). The measured fluorescence yield per atom is plotted vs the X-ray scattering efficiency for Xe clusters on the right side. Both parameters are corrected for effects from the cluster radius $r$ and X-ray wavelength $\lambda$(see Methods and Supplements for more details). The gray dashed line corresponds to tabulated Xe M-shell fluorescence yield 0.00365\cite{Hubbell1994}. Each dot in the graph represents a single cluster observed by a single FEL exposure. The point spread stems from the fact that the nanoparticle diameters 60-150\,nm are much smaller than the FEL focus average full width half maximum (FWHM) 1.5\,$\mu m$. Each randomly injected sample experiences a different exposure FEL intensity inside the FEL focus and only the top few percent of all hits with the highest fluorescence/scattering yields must have been recorded near the FEL focus center with the highest intensities.
    }
    % 700eV = 1.77nm
    % 1500eV = 0.83nm
    \label{fig:3}
\end{figure}
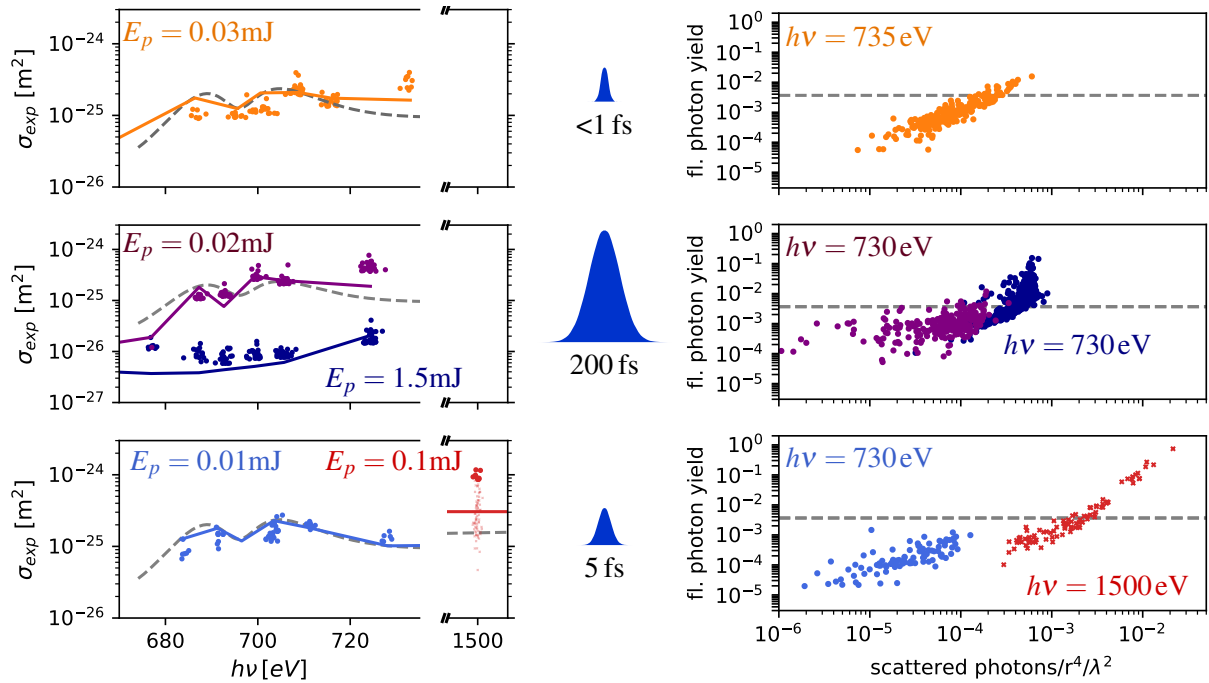

A statistical analysis of the scattering cross section $\sigma_{exp}$ reflected in thousands of recorded images supports the hypothesis that X-ray diffraction from excited ions can be both, detrimental or beneficial to the scattering cross section. This data overview illuminates the full extent of bleaching in 200-fs snapshots and the enhancement of $\sigma_{exp}$ in diffraction patterns recorded with sub-fs and few-fs pulses as summarized on the right size in \autoref{fig:3}. We calculated $\sigma_{exp}$ per Xe atom for the top 5\,\% brightest images for each incoming X-ray photon energy scan step as demonstrated in \autoref{fig:3}, left panels. Each dot represents a measured diffraction image of a single Xe nanoparticle which is sufficiently bright for an automated nanoparticle size fitting routine (see Methods and Supplemental Material). The individual value for $\sigma_{exp}$ can be directly deduced from the image brightness corrected for the exposure intensity $I_{ph}$ and the nanoparticle size. We determined the individual $I_{ph}$ based on a calibration correlating the M-shell fluorescence yield recorded parasitically by the pnCCD and the pulse energy values reported by the gas detector (see Supplements).

The corresponding literature values for $\sigma_{scat}$ for the neutral Xe atom are visualised by the dashed line \cite{Hubbell1994} (see Supplements). For all pulse durations and pulse energies below 1.5\,mJ the $\sigma_{exp}$ values follow closely the literature Xe values near the 3d resonance ($h\nu < 720\,\mathrm{eV}$). Two major tendencies deviating from the neutral Xe case emerge from the analysis. First, for all three pulse durations, conditions exist when $\sigma_{exp}$ significantly exceed the Xe literature value for the corresponding photon energy above 720\,eV. At $h\nu=1500\,$eV and 5-10\,fs pulse duration (red dots), the enhancement is 10 times higher than any point on the dotted Xe literature value curve. Second, $\sigma_{exp}$ extracted from images recorded with 200-fs and $E_{p}=1.5$\,mJ pulses (blue dots) are overall 10 times lower than Xe literature values.

A correlation between the $\sigma_{exp}$ and the number of photons absorbed per atom disentangles the complex interplay between photo-absorption, transient ion states and ultimately, structural damage of the sample. In the right panel of \autoref{fig:3} the number of scattered photons corrected for the wavelength and particles size is plotted vs the M-shell fluorescence yield per atom \suppref{Calculation of fluorescence photons per atom}. Both values are extracted from individual snapshots recorded at photon energies above 720\,eV.  The M-shell fluorescence yield per neutral Xe atom after a single photoabsorption event is only about 0.365\,\%\cite{Hubbell1994} which is marked as the dotted grey line. Higher fluorescence yield values indicate that multiple photons were absorbed. The number of scattered and absorbed photons depends on $I_{ph}$  which fluctuates mostly because of the random nanoparticle´s position within the FEL focal volume intensity distribution \cite{Bostedt2012, Gorkhover2012, Gorkhover2016, Hantke2014, Ho2020Sucrose,Ayyer2021}. Overall, the number of scattered photons increases with $I_{ph}$  in parallel with the fluorescence yield per atom and the highest yields stem from the FEL focus center.

The correlation between scattering and fluorescence reflects accelerated ionic structure damage during intense 200 fs pulses, but also non-linear increase the elastic scattering signal on short sub fs to 10 fs short time scales. First, the images, which are recorded with 200\,fs and 1.5\,mJ FEL pulses exhibit a saturation behavior where scattering efficiency stagnates with increasing fluorescence (blue dots). This is a clear signature of structural damage of the specimen as demonstrated by previous studies \cite{Gorkhover2012,Gorkhover2016, Fluckiger2016time, Peltz2022few, Ho2020Sucrose}. The scattering efficiency of the entire nanoparticle depends on the ion density which decreases dramatically during the onset of a hydrodynamic expansion . The fluorescence efficiency is less dependent on ion density and individual ions continue to absorb X-rays similarly to the ones inside the intact sample.

Second, the high fluorescence yield of of the brightest images with enhanced $\sigma_{exp}$ suggests that one or more X-ray photons per atom were absorbed during the FEL exposure. Such high ionization levels within the sample promote the role of transient form factor changes as the probability that X-rays are scattered off an ion with a core-hole rather than a neutral atom increases. This effect is very pronounced in diffraction images recorded with 1500\,eV and 5-10\,fs FEL pulses where the highest enhancement of $\sigma_{exp}$ was observed. In the brightest shots, the fluorescence yield increases by two orders of magnitude indicating very high Xe charge states during the FEL exposure.  Conversely,  $\sigma_{exp}$ is similar to neutral Xe $\sigma_{scat}$ in diffraction patterns with low levels of fluorescence recorded with weaker 5\,fs pulses at 730\,eV (light blue dots, lowerst panel).

Our simulation \cite{Ho-2017-JPB}  of the corresponding $\sigma_{exp}$ (solid line, left panels, which  is in good agreement with our measurements, attributes trends mentioned above to transient resonances \autoref{fig:resonancemaps}. For 200\,fs FEL exposures with $E_{p}=1.5$\,mJ pulses, our calculation predicts a significant lowering of the scattering cross section because of sample´s rapid disintegration partially fuelled by transient resonances \autoref{fig:clusterexpansion}. The presence of transient resonances above the 3d absorption edge for shorter and/or longer but weaker pulses is also supported by our calculation. If the sample was exposed to attenuated 200\,fs pulses, the hydrodynamic expansion occurs much slower and thus, the benefits of TR reappear. Interestingly, the simulation underestimates the non-linear scattering enhancement at $h\nu=1500$\,eV. One possible reason is that our simulation neglects ionisation potential suppression \cite{stewart1966lowering, Son2014quantum} which can rapidly delocalise the valence electron shells and lead to extremely efficient transient resonance cascades\cite{Wabnitz2002}. Assuming that Xe $n=5$ and $n=4$ shells are fully removed, our simulation predicts $\sigma_{scat}^*$ up to 9 times the neutral Xe cross section for 5-10\,fs FEL pulses at $h\nu=1500\,eV$. An indirect support for a significant role of complex nanoplasma ionization effects stems from coincident ion spectra measurement which witness extreme high and energetic ions in the brightest shots recorded with 1500\,eV pulses (see Supplements).

%%%%%%%%%%%%%%%%% Summary  and Outlook %%%%%%%%%%%%%%%%%%%%
Overall, our benchmark study illuminates the potential of the little explored transient resonances in ultrafast X-ray CDI. In essence, we demonstrate that transient resonances can significantly increase the brightness of X-ray diffraction images if FEL pulses are shorter than the onset of the ionic structure damage. We observed the enhancement effect across at least one order of magnitude in FEL intensity for pulse durations < 10 fs. This data and our simulations indicate that the effect is scalable with FEL pulse energy as subsequent photoionisation events can drive the ions into a resonant cascade similar to REXMI \cite{Rudek2012} and some resonances enhance diffraction by factors $>10^4$ (supplement \autoref{fig:resonancemaps}). The brightest images recorded here with 1500\,eV photons (0.7\,nm wavelength) reflect consistent elastic scattering levels down to 5\,nm full period spatial resolution which is to our knowledge the best result for a single exposure ultrafast soft X-ray image of a nanoscale object (\autoref{fig:experiment}, upper right panel). In principle, TRs should also exist in the hard X-ray regime which offers a direct route to sub-nm resolution. Beyond that, TRs are element specific and thus could increase material contrast in X-ray images and potentially eliminate the need of heavy stain atoms in time-resolved multi-wavelength anomalous crystallography studies.
In principle, transient resonances can be created by an optical pre-pulse and significantly increase signal-to-noise ratio in state-of-art ultrafast table-top XUV imaging setups such as \cite{Rupp2017coherent}.

Our study emphasizes the significance of intense and short FEL pulses which have received little attention so far in CDI, but experience a rapid development in accelerator science \cite{Marinelli2016,Lutman2016, guetg2018generation, Duris2020} . Newly available intense sub-fs FEL pulses can become a significant driver for ultrafast CDI as relaxation processes such as Auger decays of K- and L-shells happen mostly within hundreds of attoseconds. In our study, we have demonstrated sub-fs X-ray CDI within a single FEL exposure and also, that transient resonances can increase the brightness of images on such short time scales. Our results pave the way for a simulation guided systematic improvement of ultrafast CDI based on non-linear effects, which has the potential to combine atomic resolution with attosecond temporal precision.

%%%%%%%%%%%%%%%%%Figure 4 350 words%%%%%%%%%%%%%%%
%%%%%%%%%%%%%%%%%%%%%%%%%%%%%%%%%%Graphics %%%%%%%%%%%%%%%%%%%%%%%

\clearpage

%%%%%%%%%%%%%%%%%%%%%%%%%%%%%%Methods%%%%%%%%%%%%%%%%%%%%%%%%%%

\section*{Supplements}

\subsection*{Simulation}

We employed  Monte-Carlo/Molecular-Dynamics (MC/MD) calculations to simulate the scattering cross sections of the Xe clusters \cite{Ho-2016-PRA, Ho-2017-JPB, Ho2020Sucrose} to model the full electron and nuclear dynamics in an atomistic manner during the full duration of the X-ray pulse. In more detail, the interaction of the atom with incident XFEL pulse is treated quantum mechanically with a Monte Carlo method by tracking explicitly the time-dependent quantum transition probability between different electronic configurations. The total transition rate, $\Gamma$, between different electronic configurations $I$ and $J$ is given by
\begin{equation}
\Gamma_{I,J} = \Gamma^{P}_{I,J}+\Gamma^{A}_{I,J}+\Gamma^{F}_{I,J}+\Gamma^{RE}_{I,J}+\Gamma^{EI}_{I,J}+\Gamma^{RC}_{I,J}+\Gamma^{SE}_{I,J}.
\end{equation}
Starting from the ground state of the neutral atom, we include the contribution from photoionization $\Gamma^{P}_{I,J}$, Auger decay $\Gamma^{A}_{I,J}$, fluorescence $\Gamma^{F}_{I,J}$, resonant excitation $\Gamma^{RE}_{I,J}$, electron-impact ionization $\Gamma^{EI}_{I,J}$, electron-ion recombination $\Gamma^{RC}_{I,J}$ and stimulated emission $\Gamma^{SE}_{I,J}$.  The cross sections and rates are calculated with Hartree-Fock-Slater model \cite{Ho-2017-JPB} with relativistic corrections and spin-orbit coupling in orbital energies, following the procedure outlined in reference \cite{Herman-1963}.  Additionally, a molecular dynamics (MD) algorithm is used to propagate all particle trajectories (atoms/ions/electrons) forward in time in 1 attosecond steps. The cluster dynamics includes electromagnetic forces between the charged particles and van der waal forces among the neutral atoms.

The importance of understanding transient dynamics is that the incoming photons arriving at different times will scatter off the instantaneously populated transient states.
Similar to the treatment presented in \cite{Ho2020Sucrose}, the observed scattering response is characterized as a sum of the instantaneous scattering patterns weighted by the pulse intensity, $j_X(\tau,t)$, with FWHM duration $\tau$ and convolved with a Gaussian bandwidth profile, $g(\omega,\omega_x)$, with a central photon energy of $\omega_x$, such that
\begin{eqnarray}
  \label{eq:npSCS}
  \frac{d  \sigma}{d\Omega}(\Vec{q}) =
  \frac{d\sigma_{\textrm{th}}}{d\Omega} \frac{1}{\mathscr{F}} \int_{0}^{+\infty}\!\! d\omega \int_{-\infty}^{+\infty}\!\! dt  g(\omega,\omega_x) j_X(\tau,t) |F_{c}(\Vec{q},t)|^2,
\end{eqnarray}
where $d\sigma_{\textrm{th}}/d\Omega$ is the Thomson scattering cross section.
\begin{equation}
    \mathscr{F} = \int_{0}^{+\infty}\!\! d\omega \int_{-\infty}^{+\infty}\!\! dt  j_X(\tau, t) g(\omega,\omega_x)
\end{equation}
is the fluence of an XFEL pulse, and $\int_{0}^{+\infty}\!\! d\omega(\omega,\omega_x) = 1$ . Here $F_{c}(\Vec{q},t)$ is the time-dependent form factor of the target cluster and is given by
\begin{equation}
  \label{eq:npFormFactor}
 F_{c}(\Vec{q},t)=\sum_{j=1}^{N_a} f_j(\Vec{q},C_j(t))e^{i \Vec{q} \cdot \Vec{R}_j(t)} + \sum_{j=1}^{N_e(t)} e^{i \Vec{q} \cdot \Vec{r}_j(t)}\, ,
\end{equation}
where $N_a$ is the total number of atoms/ions, $\Vec{R}_j(t)$, $C_j(t)$ and $f_j(\Vec{q},C_j(t))$ are the position, the electronic configuration and the atomic form factor of the $j$-th atom/ion respectively.  $N_e(t)$ is the number of delocalized electrons within the focal region of the X-ray pulse and $\Vec{r}_j(t)$ are their positions.

To simulate the scattering cross section of atomic xenon, we used the Monte-Carlo rate equation \cite{Ho-2014-PRL}.  We explicitly track the time-dependent quantum transition probability between different electronic configurations in atoms exposed to X-ray pulse. The total transition rate, $\Gamma$, between different electronic configurations $I$ and $J$ is same as that used in MC/MD calculation, except $\Gamma^{EI}_{I,J}$ and $\Gamma^{RC}_{I,J}$ are set to zero.

\subsection*{Simulation Results}

\begin{figure}
    \centering
    \includegraphics[width=7cm]{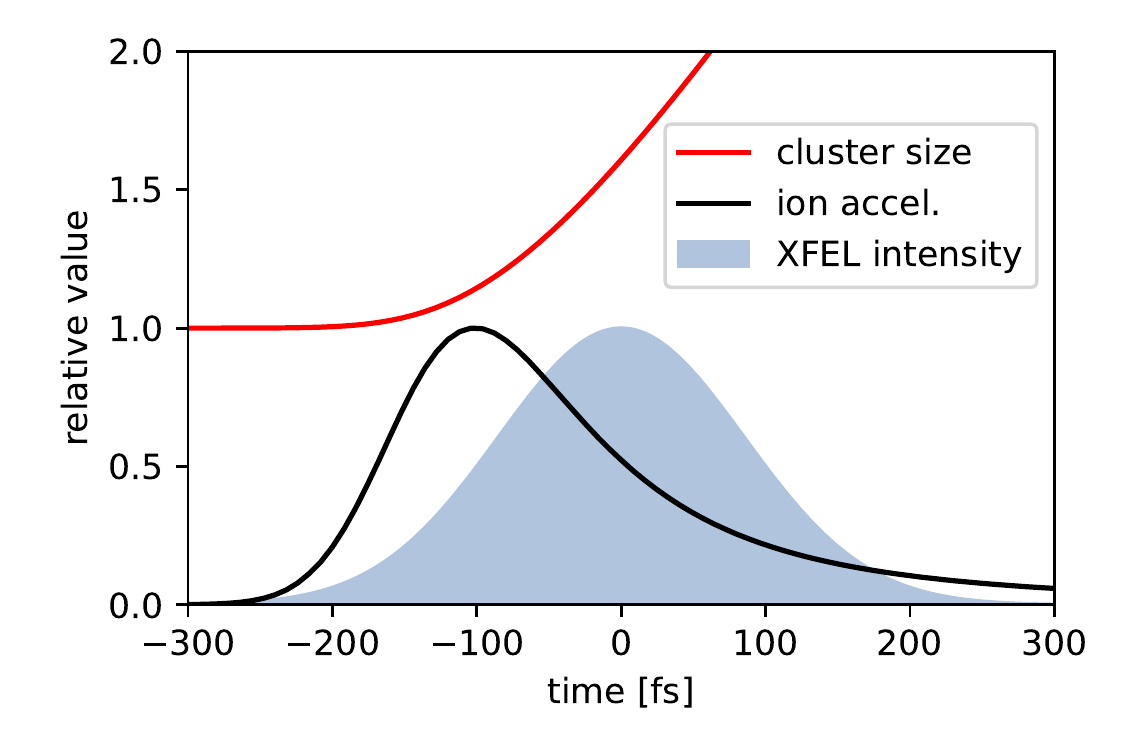}
    \caption{Simulation of Xe cluster expansion due to XFEL irradiation with a 200\,fs pulse: The cluster's size is measured by the standard deviation of all particle positions within the simulation and the \enquote{ion acceleration} is its second derivative. The acceleration peaks around $-100$\,fs, where the ionization rate also has a maximum due to collisional ionization (not shown).}
    \label{fig:clusterexpansion}
\end{figure}

At the beginning of the simulation, neutral atoms interact through a Lennard Jones potential.
X-ray photoionisation and Auger processes produce ions and free electrons, adding Coulombic forces to the cluster.
These electrons in turn trigger collisional ionization which rapidly becomes the dominant ionization process. The delocalized but trapped electrons heat up the cluster from the inside.
This process peaks around $-100$\,fs together with the acceleration of the xenon ions as seen in \autoref{fig:clusterexpansion}.
In turn the cluster expands well before the maximum of the XFEL pulse arrives.
The simulation results in \autoref{fig:3} are based on the same method and additionally integrated the scattering cross section over all xenon atoms and the entire exposure time.

\begin{figure}
    \centering
    \includegraphics[width=10cm]{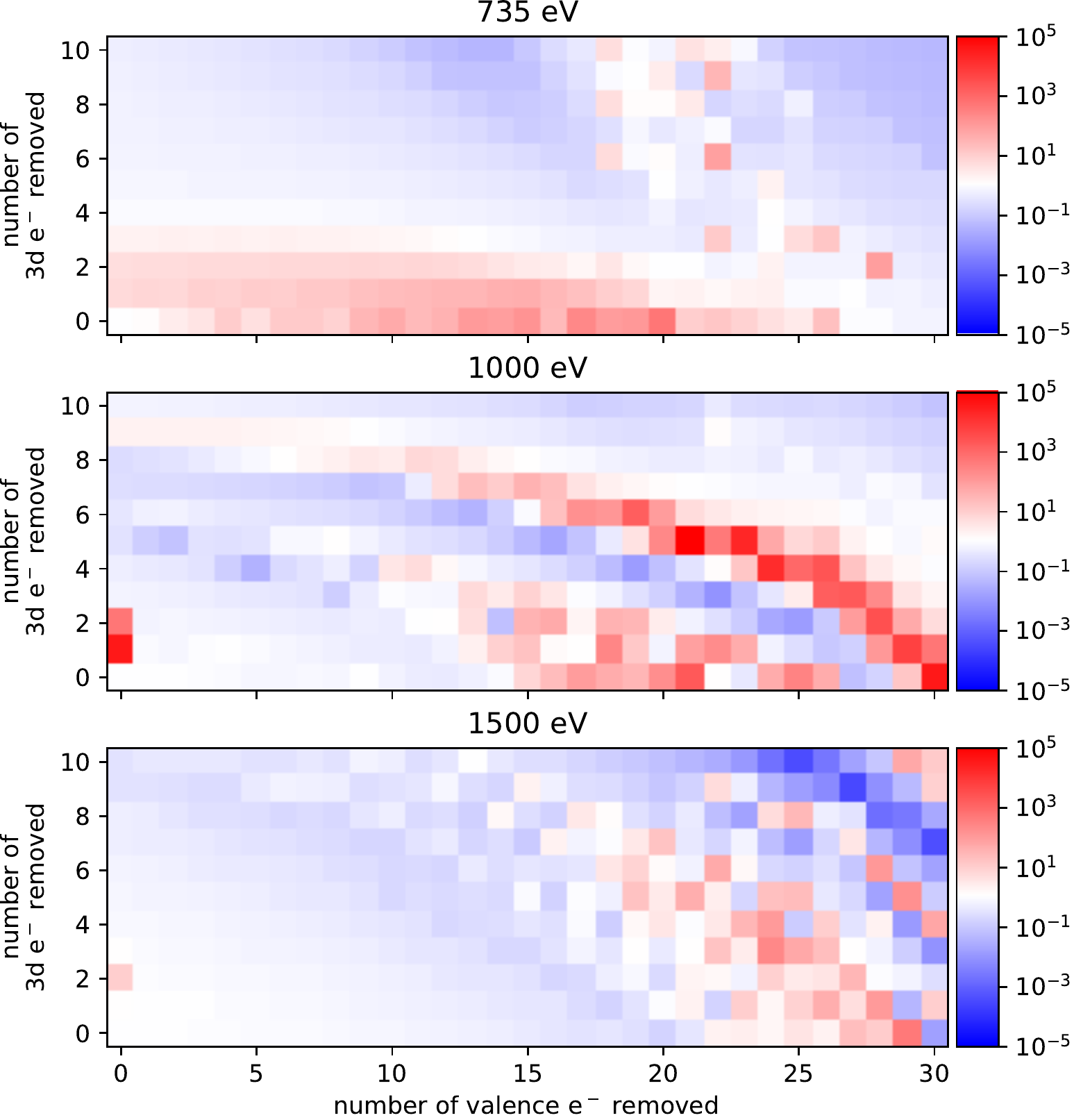}
    \caption{Resonance maps for specific electronic states of Xenon: For a number of removed valence electrons (x-axis) and a number of removed inner shell (3d) electrons (y-axis), the colorscale shows the scattering cross section relative to the neutral Xe cross section. At a photon energy $h\nu=735\,\mathrm{eV}$ enhancement already sets in for low charge states, while at 1500\,eV higher charge states are required for enhancement. The simulations suggest that a single charge state can scatter up to a factor $10^4$ more than the neutral Xenon at 1500\,eV. At 1000\,eV an enhancement of up to $10^5$ has been calculated for specific charge states.}
    \label{fig:resonancemaps}
\end{figure}

We also simulated the ionic scattering cross sections for individual electronic configurations of the Xe ions (\autoref{fig:resonancemaps}).
For simplicity, an electronic configuration is represented by the number of removed valence electrons (x-axis) and the number of removed inner shell (3d) electrons (y-axis). The colorscale displays the change of the scattering cross section relative to neutral Xe. Some configurations are exceeding factor $10^5$ enhancement of the scattering cross section compared to neutral Xenon. In some regions of \autoref{fig:resonancemaps}, clustering of multiple strong TRs indicate that the system can stay resonant during multiple absorption events.
At an incident photon energy $h\nu=735\,\mathrm{eV}$, enhancement is already possible for relatively low charge states below 15.
At $h\nu=1500\,\mathrm{eV}$, very high charge states around 25 are required and show significant enhancement of $\sigma^*$ of up to factor $10^4$.
Those very high ion charge states are consistent with the measurement in our experiment.
A photon energy of 1000\,eV shows the greatest enhancement.
In future research, such calculations can be used to identify and target specific electronic configurations and FEL pulse parameters. In the ideal case, one would identify regions with multiple TRs which form a cascade towards higher charge states. This way brightness enhancement can be scaled in combination with high peak power densities and short pulses.

\subsection*{Experimental Setup}

\begin{figure}
  \centering
  \includegraphics[width=12cm]{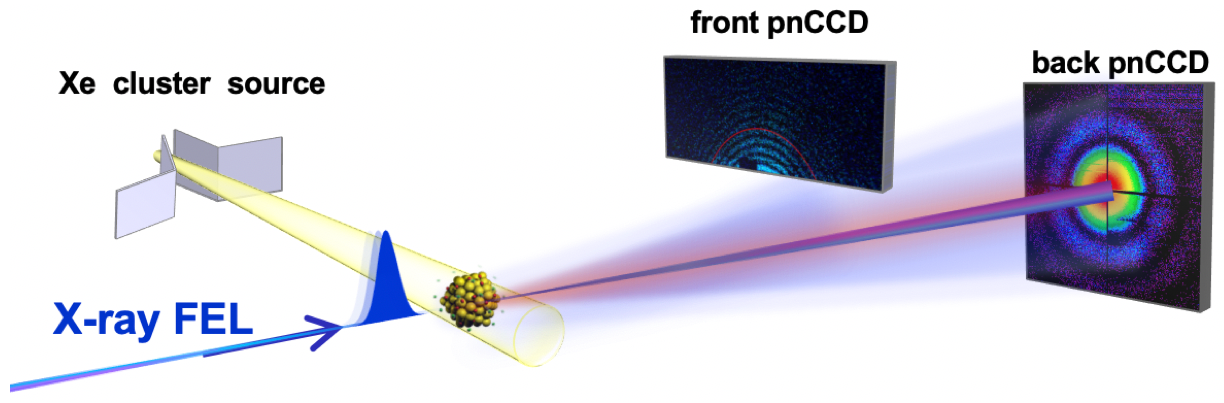}
  \caption{Schematic of the experimental setup at the AMO Hutch at LCLS: Intense X-ray FEL pulses were focused and intersected with a stream of Xe clusters. The resulting single- shot single-particle X-ray diffraction patterns were recorded using pnCCD detectors positioned further downstream of the interaction point. The front pnccd is positioned in a distance of 130\,mm and the rear pnccd at 732\,mm from the X-ray focus. The outer regions of the front pnccd, where the diffraction signal has vanished still register single fluorescence photons.}
  \label{fig:setup}
\end{figure}

A schematic of our experiment is exhibited in \autoref{fig:setup}: Individual Xe nanoparticles with diameters 40-150 nanometers intersect the path of the focused and intense single X-ray FEL pulses at the LAMP endstation at the Linac Coherent Light Source (LCLS) \cite{Ferguson2015}. Single particle, single exposure snapshots were recorded using a p-n junction charge coupled devices (pnCCDs)\cite{Strueder2010} located at two different positions further downstream from the interaction region. The particle size is encoded with $\pm 0.3$\,nm precision into the Airy pattern-like diffraction \suppref{Cluster size fitting model}. Single-exposure radial diffraction profiles collected by the rear and front detectors are in agreement with theoretical plots for X-ray diffraction by a perfect sphere within Born approximation \cite{Guinier1955}. The deviations from a perfect sphere can be mostly attributed to sharp facets near the surface.

The experiment was conducted at the AMO hutch of the LCLS light source at SLAC National Laboratory in the US. Xenon nanoclusters were injected into the focal region of the X-ray beam (see \autoref{fig:setup}). A pnCCD detector was positioned 130\,mm downstream of the interaction point thus called the \enquote{front} pnCCD Another, the \enquote{rear} pnccd, was positioned at a distance of 732\,mm downstream the interaction point. The Xe cluster density was adjusted to record diffraction patterns of single Xe nanoclusters only. As those clusters are almost perfect spheres their size can be reconstructed with high precision. The number of scattered photons on the detector and the exact size of the cluster combined allow to calculate the scattering cross section of a single Xe atom. See \suppref{Cluster size fitting model} for details. This step is of crucial importance for the analysis as it allows us to correct the scattering for the cluster size in each XFEL pulse and thus calculate the scattering cross section.

The XFEL focal spot of  1.5\,$\mu$m (FWHM) is over one order of magnitude wider than the nanoparticles and thus, each nanoparticle is exposed to a different but uniform power density. Based on statistical considerations one can assume that brightest 0.1\,\% of all shots were recorded close to the FEL focus center. The absolute exposure power density in the focus center has been calibrated by independently measured Argon atomic spectra and cross checked with beam line characteristics.

\begin{table}[h]
    \centering
    \begin{tabular}{p{1.5cm}p{1.7cm}p{0.7cm}p{1.2cm}p{2cm}p{1.9cm}}
    name & color & {$\tau$ \newline [fs]} & Energy \newline [mJ] & {Fluence \newline [$\mu \mathrm{J}/\mu \mathrm{m}^2$]} & Photon \newline Energy \\ \hline

    XLEAP & \textcolor{orange!90!black}{orange} & <1 & 0.03 & $0.2\pm 0.1$  & Xe M-edge \\
    200fsfull & \textcolor{blue!50!black}{dark blue} & 200 & 1.5 & $12\pm 2$ & Xe M-edge \\
    200fsatt & \textcolor{purple!50!black}{purple} & 200 & 0.02 & $0.16\pm 0.02$  & Xe M-edge \\
    5fs & \textcolor{royalblue!90!black}{light blue} & 5 & 0.01 & $0.08\pm 0.03$ & Xe M-edge \\
    1500\,eV & \textcolor{red!80!black}{red} & 5 & 0.1 & $0.8\pm 0.3$ & 1500\,eV

    \end{tabular}
    \caption{Summarized experimental conditions with color coding used throughout the publication. The fluence is shown in average values over each dataset and $\pm$ values indicate the stability at a single photon energy. The underlying data is the pulse energy measured by the gas detectors.}
    \label{tab:colors}
\end{table}

While recording these diffraction images, we systematically scanned the photon energy around the Xenon M-shell absorption edge ($\sim$700\,eV).
Three different pulse durations were used throughout the experiment: \\
1) 200\,fs duration, \\
2) 5\,fs pulse duration by inserting a slotted foil\cite{osti_833050} and \\
3) the recently developed XLEAP\cite{Duris2020} mode delivering sub-fs pulses. \\
For each setting, the attenuation was adjusted to keep the photon number constant in each X-ray pulse, all the parameters are summarized in \autoref{tab:colors}
Inside the high-vacuum chamber the X-ray beam was focused down to a size of $\sim 1.5\mu m$ (FWHM) in diameter using the existing KB-mirrors of the AMO endstation. The beamline transmission was measured to be 0.2 around 700\,eV and 0.4 around 1500\,eV.

The Xe clusters were formed by supersonic expansion of xenon at 9 bar backing pressurethrough a 200\,$\mu$m conical nozzle with a half-opening angle of 4$^\circ$ cooled to 245\,K. The pulsed cluster jet was skimmed twice. An additional piezo-driven slit skimmer right before the interaction region was adjusted such that, on average, less than one particle was in the FEL focus.

\subsection*{Cluster size fitting model} %\label{sec:clustersize}

We analyzed single diffraction snapshots of individual clusters and  fitted the corresponding diffraction radial profile depending on the cluster size. As clusters are mostly spherical, we choose the analytical diffraction pattern of a hard sphere as our working model \cite{Guinier1955}. The adjustable parameters of the aforementioned  fit determine the particle size (namely its radius $R$) and the brightness of the recorded diffraction image encoded in the form of the parameter $I_0$. We extracted the absolute scattering cross section from a single image by combining both fit parameters with the calibrated X-ray exposure intensity values (last section, Supplements).

\begin{figure}
    \centering
    \includegraphics[width=7cm]{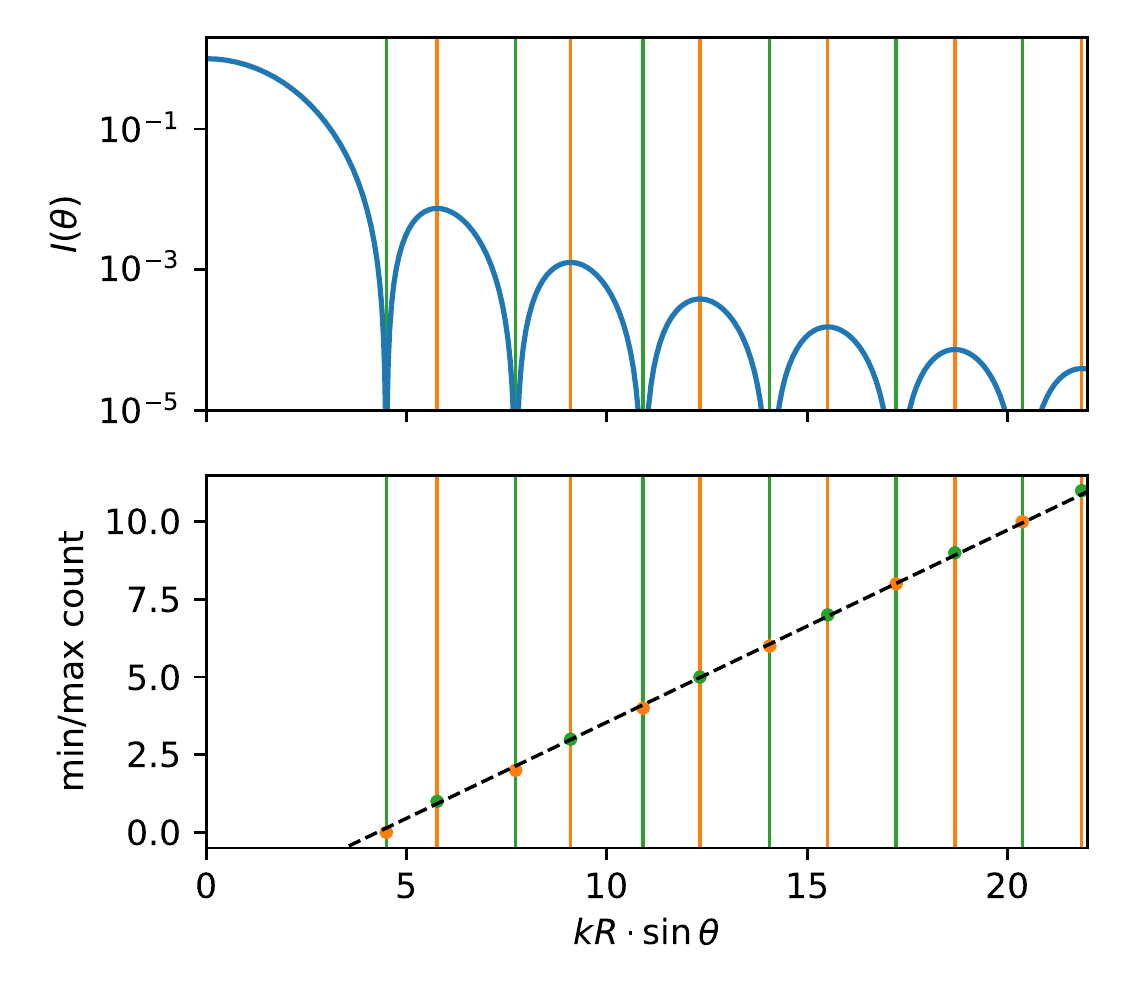}
    \caption{The diffraction pattern of a solid sphere $I(\theta)$ is shown on the top. The green (orange) lines mark the minima (maxima) of the this function. The fact that maxima and minima are almost equally spaced is used to estimate their spacing and thus estimate the particle radius $R$. This estimated value is then used as a starting value for fitting the analytical diffraction pattern $I(\theta)$ to the measured one and thus determine $R$ as well as $I_0$. This is performed separately for every measured single shot diffraction pattern in the experiment.}
    \label{fig:diffaprox}
\end{figure}

The one-dimensional plot of the X-ray diffraction patterns from a solid sphere is given by and displayed in \autoref{fig:diffaprox} (top, blue line):
\begin{align}
I(\theta) = I_0\cdot\frac{9}{x^2} \left(\frac{\sin(x)}{x^2} - \frac{\cos(x)}{x}\right)^2
\end{align}
with $x=kR\sin\theta = \frac{2\pi}{\lambda}\cdot R \cdot \sin\theta$
and the sphere's radius $R$. The fitting route starts with a reasonable estimate for $I_0$ and $R$. This is particularly important for the parameter $R$, as the fitting routine may otherwise converge to a local minimum. For the numerical calculations, it is much more practical to write $x=kR\sin\theta$ as $x=4.493 \cdot \frac{x_\text{pxl}}{x_1}$ where $x_\text{pxl}$ is the x-value in pixel on the detector and $x_1$ is the pixel number of the first minimum of the function $I(\theta)$. The number 4.493 is the position of the first minimum of $I(\theta)$ as seen in \autoref{fig:diffaprox} and was determined numerically.

Any given individually measured single shot diffraction image is first preprocessed by masking invalid detector regions and over-exposed pixels. A radial profile of the remaining image is calculated, which represents the measured $I(\theta)$. A peak finder will mark the local maxima and minima, which corresponds to the green and orange positions in \autoref{fig:diffaprox}. As long as the diffraction pattern is clearly visible, maxima and minima will be alternating. As soon as the signal vanishes in noise, however, the peak finding will fail to identify the peak positions correctly and multiple maxima or minima will directly follow each other. At that point any further peaks will be deemed invalid. A straight line will be fitted to all valid peak positions (black dashed in \autoref{fig:diffaprox}). The slope of that line is characteristic for the distance between minima and maxima and thus characteristic for the size $R$ of the particle. The slope is used to calculate the estimate for $R$, that is required as a starting value to fit $I(\theta)$ to the measured $I(\theta)$. Please note that a bias of this estimate is tolerable as the fit will converge towards the measured shape of $I(\theta)$.

In a subsequent step the total number of scattered photons $N_\text{scat}$ is calculated from the values $R$ and $I_0$ by integration over the diffraction pattern.

\subsection*{Calculation of the scattering cross section}

Combining the estimated fluence and the parameters $R$ and $N_\text{scat}$ (calculated from $R$ and $I_0$, see cluster size fitting model) allows us to calculate the scattering cross section of the Xe nanocluster $\sigma_\text{sphere}$ and, subsequently, the scattering cross section of an individual Xe atom $\sigma_\text{scat}$.

Scattering of an individual atom is given by\cite{Attwood2012}
\begin{align}
\sigma_\text{scat} = \frac{8}{3}\pi r_e^2 |f|^2
\end{align}
A nanocluster consisting of multiple atoms has a much larger scattering which is given by the coherent sum of all contributing atoms (Eq. 30\cite{Howellst1995}):
\begin{align}
\sigma_\text{sphere} &= 2\pi r_e^2 \lambda^2 R^4 \rho^2 |(f_1 + if_2)|^2 \\
 &= \underbrace{\frac{9}{16\pi} \lambda^2 R \rho}_{=:X_\text{sphere}} \cdot N_\text{atoms} \cdot \sigma_\text{scat}
\end{align}
where $r_e$ is the classical electron radius, $\lambda$ the incident wavelength, $R$ the radius of the nanocluster, $\rho$ the particle density of the nanocluster and $f_1$ and $f_2$ are the real and imaginary part of the atomic scattering factor.
For simplicity, the terms are rearranged such that $X_\text{sphere}$ is the factor by which a nanocluster scatters more than the incoherent sum of all its atoms.

Now, the total radiated energy $E_\text{scatt}$ can be related to the incident fluence $F$. By definition
% Attwood, Eq 2.38 (p.42)
\begin{align}
    E_\text{scat} = \sigma_\text{sphere} \cdot F
\end{align}
and the atomic scattering cross section $\sigma_\text{scat}$ can be calculated from the measurement via
\begin{align}
    \sigma_\text{scat} = \frac{E_\text{scat}}{N_\text{atoms} X_\text{sphere} F}
\end{align}
with $N_\text{atoms}$ as the total number of Xe atoms in the nanocluster. This relation is used in the data analysis to calculate the atomic scattering cross section of Xe. The number of Xe atoms in the cluster $N_\text{atoms}$ is calculated by multiplying the volume of the cluster and the number density of solid Xenon:  $N_\text{atoms} = \rho_\text{Xe} \frac{4}{3} \pi R^3$ with $\rho_\text{Xe} = 1.78\cdot 10^{28} \frac{\text{atoms}}{{m^3}}$ \cite{Klein1977}

(footnote: In the 5\,fs 3d edge run we could in principle estimate the FEL exposure power density for each hit based on the fluorescence yield, but this may be misleading out of consistency considerations as other runs are more dominated by TRs and the fluorescence yield changes substantially)

\subsection*{Single exposure images}
\begin{figure}
    \centering
    \includegraphics[width=8cm]{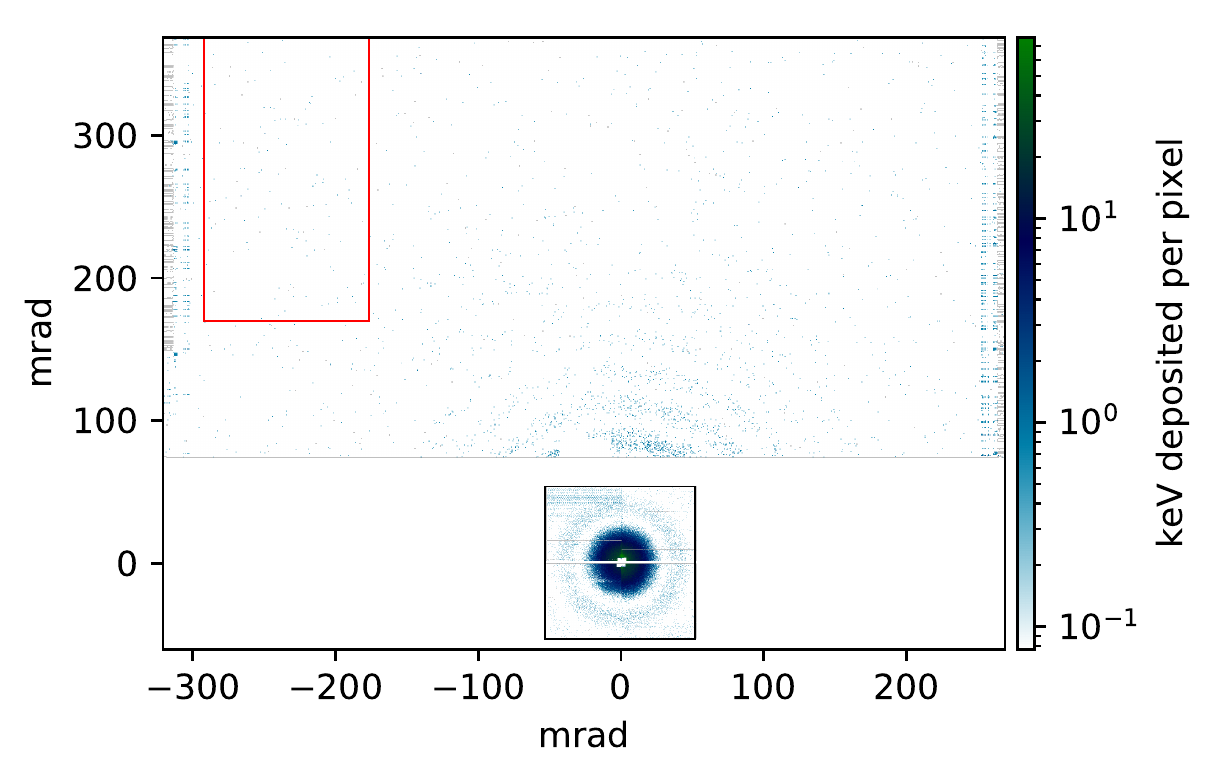} \\
    \includegraphics[width=8cm]{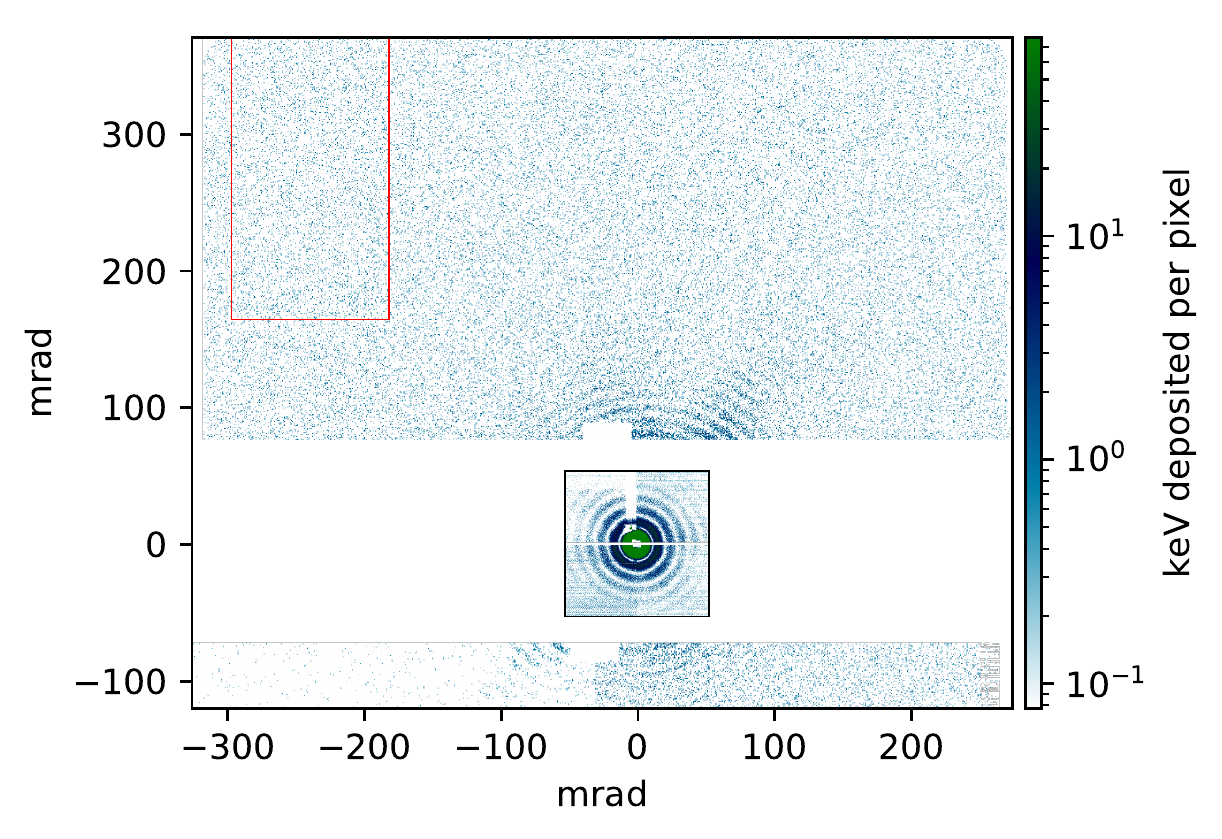}
    \includegraphics[width=8cm]{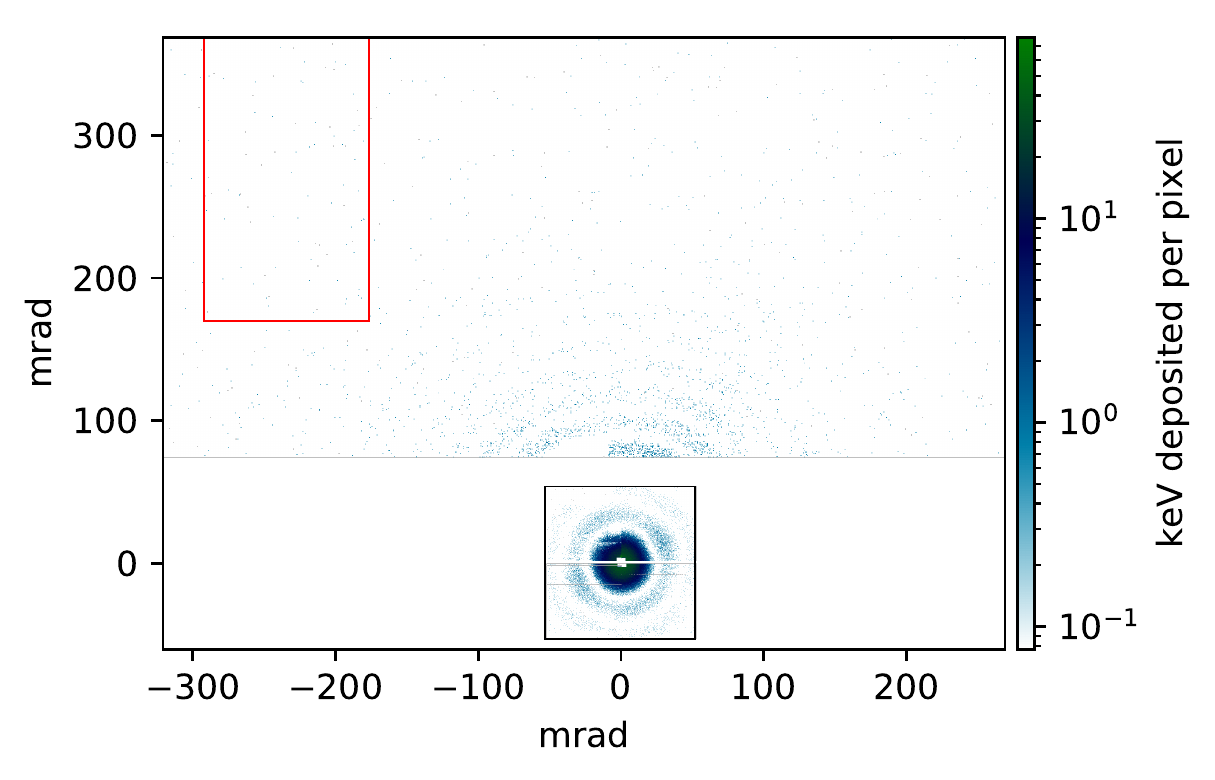}
    \caption{Single shot images for the same conditions as shown in \autoref{fig:experiment}: The top image is for the 200\,fs pulse, lower left was taken at 5\,fs at 1500\,eV and lower right image is for XLEAP conditions (<0.5\,fs). The image combines the data of the front and the rear pnCCD. The single photons are clearly visible in the images. Please note, that a single photon appears smaller on the rear detector compared to the front detector as each pixel spans a much smaller solid angle. All three particles are of very similar size. The lower left image shows faster diffraction rings because the incident photon energy is 1500\,eV compared to $\sim$700\,eV for the other two. The red rectangle indicates the region which has been used to evaluate the fluorescence yield.}
    \label{fig:singleshots}
\end{figure}

\autoref{fig:singleshots} shows the detector images corresponding to radial profiles shown in \autoref{fig:experiment}. The images are combined from the rear pnccd around the beam axis, covering angles from -60 to 60\,mrad and the front pnccd half detector covering angles from 110\,mrad and beyond. In the experiment (compare \autoref{fig:experiment}a) the rear pnccd was a factor 5.7 further away from the xray focus than the front pnccd. At all experimental conditions the diffraction image of the Xe cluster clearly extends to the region covered by the front pnccd. The clusters shown here are of almost identical size. First note, that and the diffraction images taken with 200\,fs and XLEAP (<0.5\,fs) look almost identical although the 200\,fs pulse carried about 60 times more energy! At the lower left image (1500\,eV) the front pnccd was adjusted slightly further from the beam axis resulting in a larger gap between the detectors, visible in \autoref{fig:singleshots} and \autoref{fig:experiment}c.

Further, the high fluorescence levels are clearly visible at the lower left image. It was taken at 1500\,eV photon energy, which is why the diffraction ring spacing changed, while the cluster size stayed the same. The region marked by the red box was used to evaluate the fluorescence yield, see \suppref{Calculation of fluorescence photons per atom}.

\subsection*{Calculation of fluorescence photons per atom}

In contrast to previous studies, which relied on focal averaged Ar gas measurements \cite{Duris2020}, we exploit the fluorescence yield of single moderately bright images for absolute X-ray exposure intensity calibration.
The comparison between the fluorescence yield from the moderate 5\,fs measurement (light blue dots) and the M-shell fluorescence yield literature value pointed out as a dashed grey line in \autoref{fig:3} a) suggests that in this run only up to about 0.3-0.4 X-ray photons/atom were absorbed.
TRs play little role if the majority of atoms remain neutral during the FEL exposure and thus, the absolute fluorescence yield is proportional to the exposure FEL power density.
Fluorescence photons were detected as single photon hits on the pnCCD detectors.
The fluorescence photons per atom (shown in \autoref{fig:3}) are calculated from an outer region of the front pnCCD by summing up all pixels above a threshold corresponding to 125eV.
This outer region (marked as a red rectangle in \autoref{fig:singleshots}) is more than 300\,mrad from the beam axis, ensuring that even the brightest diffraction patterns have no signal at those large angels and consequently only the fluorescence is measured.
The energy deposited on the detector corner is then multiplied by a factor to scale this number to the energy emitted into $4\pi$ assuming isotropically emitted fluorescence and a correction for the quantum efficiency of the pnccds.
It is further divided by the number of atoms in the cluster, which finally yields the “fluorescence photons per atom” as shown in \autoref{fig:3}.

\subsection*{Calibration of incoming X-ray photon energy}

\begin{figure}
    \centering
    % use png. pdf has too many points and slows the pdf down.
    \includegraphics[width=8cm]{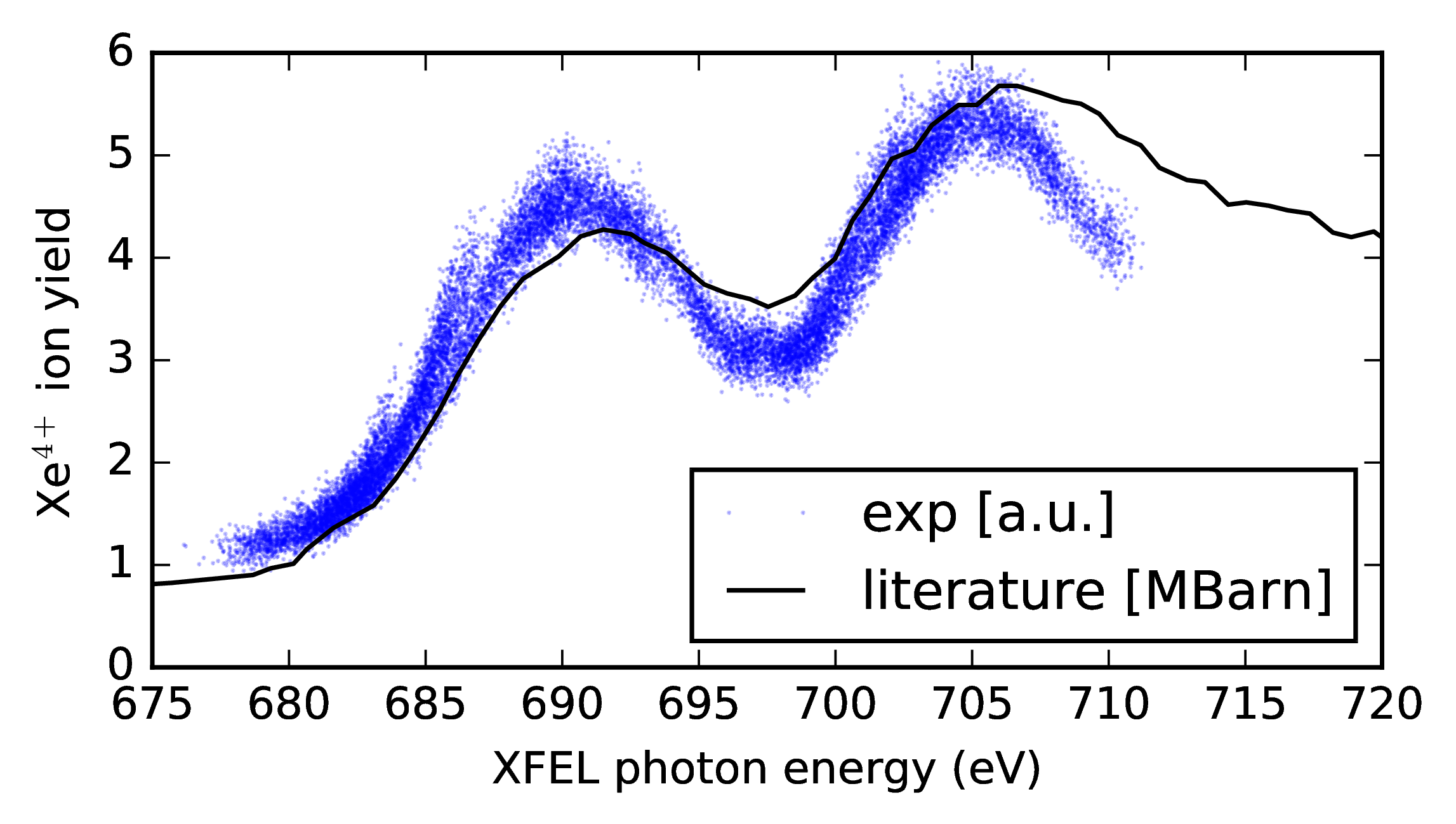}
   \caption{Calibration of the incoming photon energy: the literature values (black) are taken from \cite{Deslattes1968}. The peak positions of our measurement and the literature values are only 0.8\,eV apart. This difference was therefore not correct in any other data shown in this publication.}
    \label{fig:calibincoming}
\end{figure}

An time of flight (tof) spectrum of gaseous Xe was measured, in order to calibrate to an absolute photon energy. \autoref{fig:calibincoming} shows  the Xe$^{4+}$ ion yield for different photon energies and compares it to a literature absorption measurement\cite{Deslattes1968}. The double peak structure is a result of the resonance ionization of the Xe 3d$_{\frac{3}{2}}$ and 3d$_{\frac{5}{2}}$ energy levels.
The comparison of the peaks positions with the literature values displays a shift of the incoming XFEL photon energy by only 0.8\,eV and was therefore not compensated in the data.

\pagebreak[0]

\section*{Acknowledgements}

Use of the Linac Coherent Light Source (LCLS), SLAC National Accelerator Laboratory, is supported by the U.S. Department of Energy (DOE), Office of Science, Office of Basic Energy Sciences (BES) under Contract No. DE-AC02-76SF00515. S.K. and T.G. were supported by the U.S. DOE BES Chemical Sciences, Geosciences and Biosciences Division through the Panofsky fellowship from SLAC National Laboratory. PJH and LY were supported by the U.S. DOE BES Chemical Sciences, Geosciences and Biosciences Division under Contract No. DE-AC02- 06CH11357. F.Z.and M.R.W were supported by the U.S. DOE BES Chemical Sciences, Geosciences and Biosciences Division,  Chemical Sciences, Geosciences and Biosciences Division through the Early-Career Research Program project number 100482.

\section*{Author contributions statement}
T.G. and A.M conceived the idea for the experiment based on discussions with C.B., J.D., J.P.MA., A.L., and A.M. developed and operated the XLEAP sub-fs mode. M.-F.L., X.L., K.N., J.W.A., and P.W. prepared the beam line for the experiment. The experimental setup was planned and performed by all authors. S.K. lead the data analysis. P.J.H performed the simulation of the experimental observations. S.K, P.J.H., and T.G. wrote the manuscript with input from all authors. All authors reviewed the manuscript.

\section*{Additional information}
The authors declare no competing interests.

\bibliography{biblio.bib}

\end{document}